\documentclass{caosp306}

\usepackage{graphicx}
\usepackage{subcaption}

\usepackage{natbib}
\articleNo{1}
\pubyear{2019}
\volume{1}
\volnumber{1}
\firstpage{1}
\received{November 11, 2019}
\accepted{}

\def\BibTeX{{\rm B\kern-.05em{\sc i\kern-.025em b}\kern-.08em
             T\kern-.1667em\lower.7ex\hbox{E}\kern-.125emX}}

\begin{document}

%
\hauthor{David V. Martin}

\title{Circumbinary Planets -- The Next Steps}


%
\author{
        David~V.~Martin \inst{1,} \inst{2,} \inst{3}}

%
\institute{
           University of Chicago, Department of Astronomy \& Astrophysics, 5640 S Ellis Ave, Chicago, IL-60637, USA 
           \and 
           The Ohio State University, Department of Astronomy, 100 W 18th Ave, Columbus, OH-43210, USA \email{martin.4096@osu.edu}
         \and 
           Fellow of the Swiss National Science Foundation
          }


\maketitle

\begin{abstract}
The Kepler mission opened the door to a small but bonafide sample of circumbinary planets. Some initial trends have been identified and used to challenge our theories of planet and binary formation. However, the Kepler sample is not only small but contains biases. I will present a circumbinary plan for the future. Specifically, I will cover the BEBOP radial velocity survey, the latest TESS transit mission and a new technique for digging out small circumbinary planets in archival Kepler photometry.
\end{abstract}

%
\section{Introduction}\label{sec:intro}
Binary stars are common. Exoplanets are common. It is natural to seek planets in binaries. Planets in binary star systems come in two flavours: circumbinary planets on exterior orbits around tight binaries, and circumstellar planets on interior orbits around one of the two components of a wide binary. Here we will only consider circumbinary planets. The discovery of Kepler-16 \citep{doyle2011} really kicked off a search which had been anticipated since before the dawn of exoplanet discoveries \citep{borucki1984,schneider1991}. A dozen or so transiting circumbinary planets have been found by this mission (reviews in \citealt{welsh2018,martin2018}), but this paper will look beyond the existing Kepler discoveries.

\section{Trends And Open Questions In Circumbinary Planets}

The dozen circumbinary planets discovered to date exhibit a few interesting trends and pose a few interesting questions.

 \begin{figure*}  
\begin{center}  
	\begin{subfigure}[b]{0.99\textwidth}
		\caption{}
		\vspace{-0.5em}
		\includegraphics[width=\textwidth]{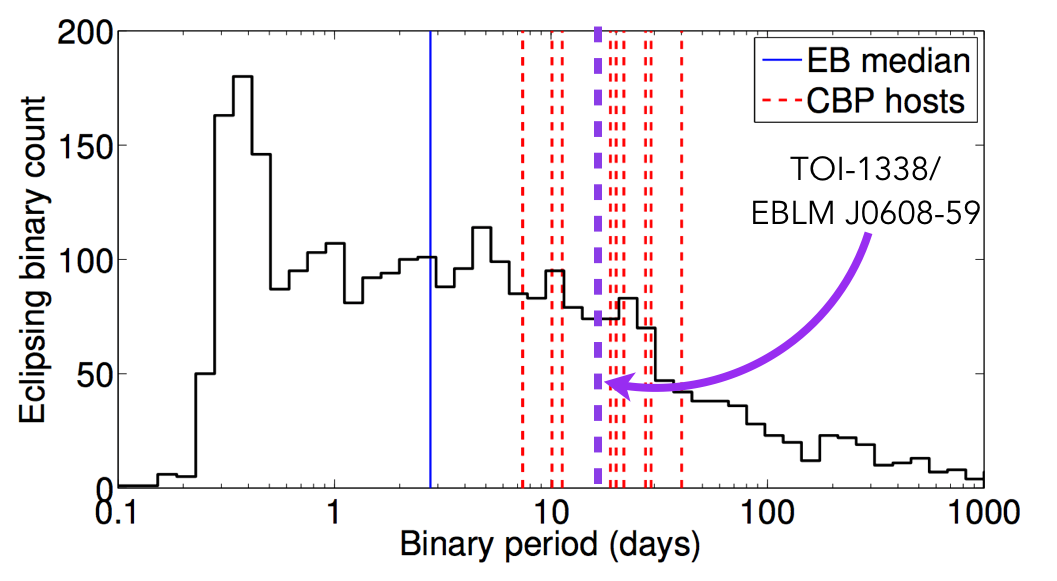}  
		\label{fig:EB_histogram}  
	\vspace{-1.5em}
	\end{subfigure}
	\begin{subfigure}[b]{0.99\textwidth}
		\caption{}
		\vspace{-0.5em}
		\includegraphics[width=\textwidth]{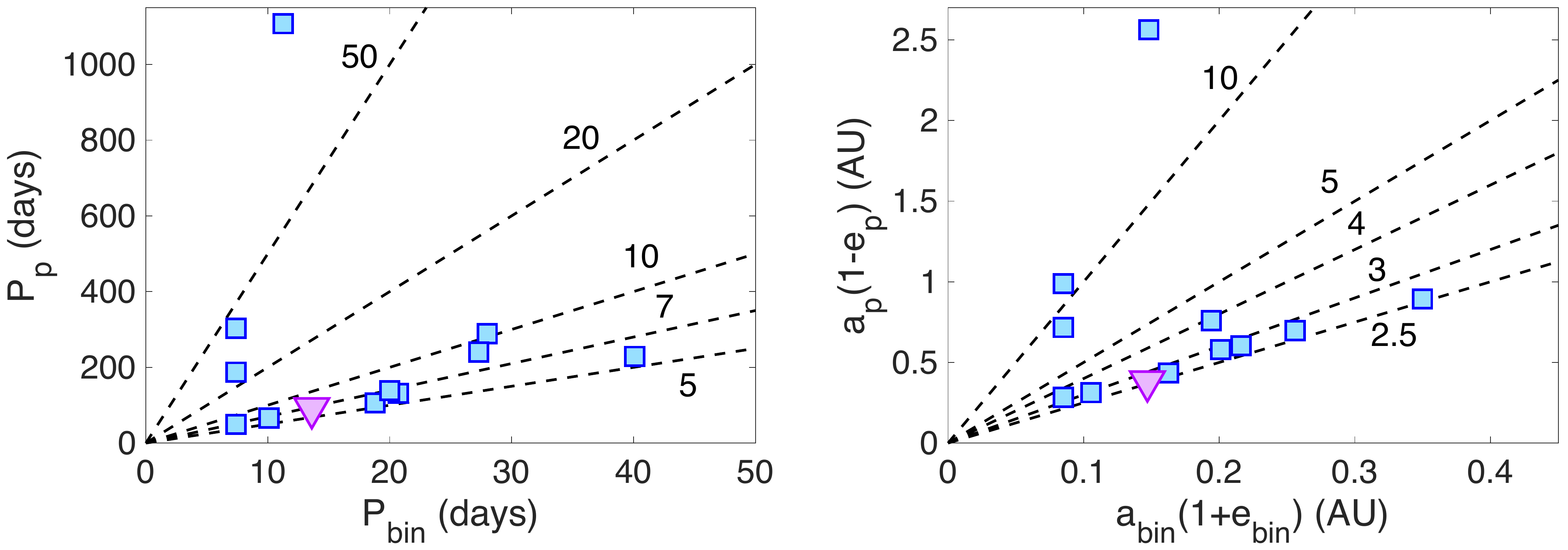}  
		\label{fig:period_semi_ratios} 
	\vspace{-1.5em} 
	\end{subfigure}	
	\begin{subfigure}[b]{0.99\textwidth}
		\caption{}
		\vspace{-0.5em}
		\includegraphics[width=\textwidth]{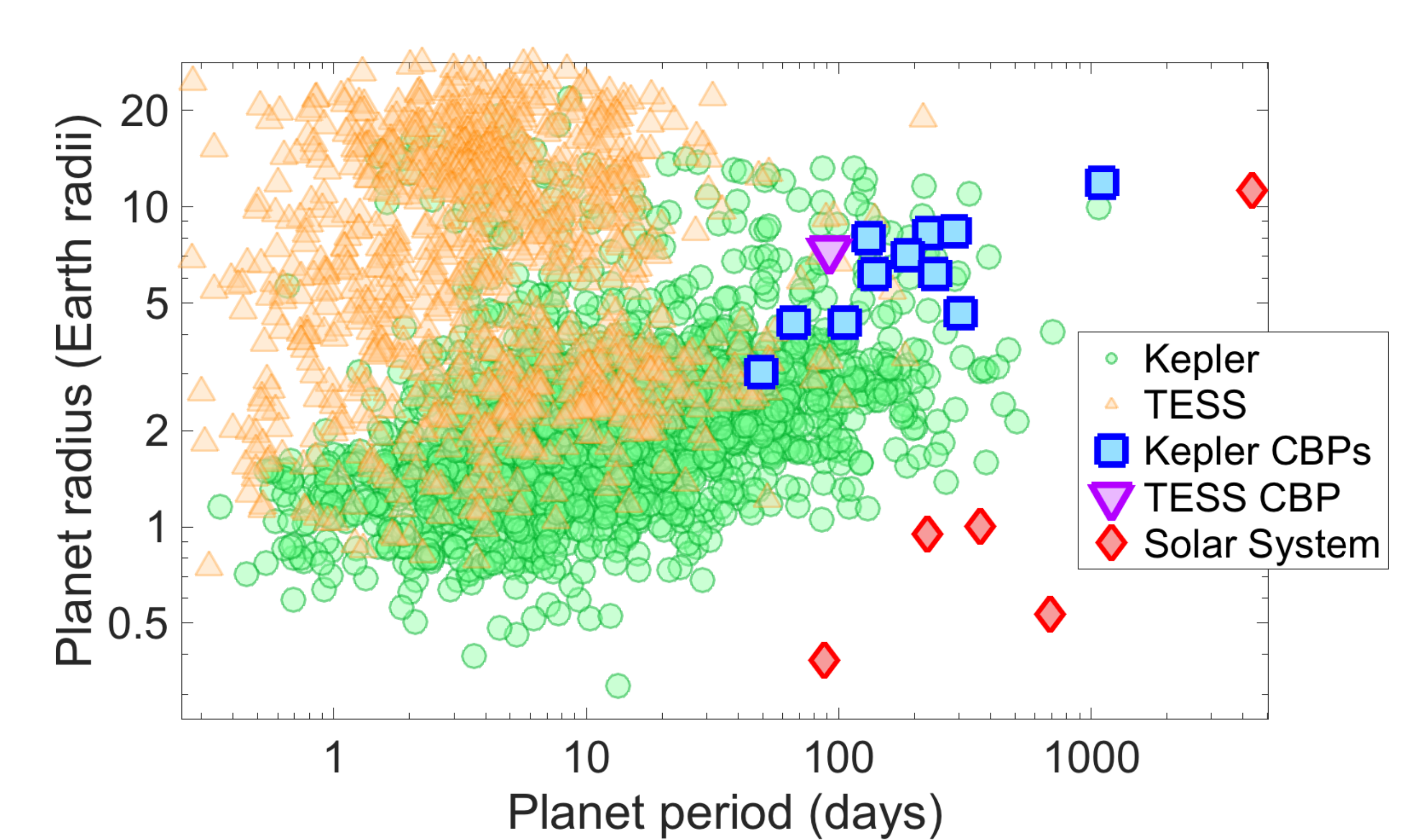}  
		\label{fig:period_radius}  
	\vspace{-1.5em}
	\end{subfigure}
	\caption{a) Histogram of Kepler eclipsing binary periods, compared with the periods of binaries known to host circumbinary planets, and the first TESS discovery of TOI-1338/EBLM J0608-59. b) Left: planet and binary periods, Right: planet periapse and binary apoapse. There is a tend for common ratios, which place the planets close to the stability boundary \citep{holman1999}. c) Size and period of Kepler and TESS discoveries around both single and multiple stars. Figures reproduced from Kostov et al. (under review) and \citet{martin2018}.}

\label{fig:trends}
\end{center}  
\end{figure*} 

\begin{enumerate}
\item \textbf{\textit {There is a dearth of circumbinary planets orbiting the tightest eclipsing binaries (EBs). }} Most EBs have a very short period ($\sim 2-3$ days) but the transiting planets are only around $>7$ day binaries (Fig.~\ref{fig:EB_histogram}). \citet{munoz2015,martin2015a,hamers2016} explained this dearth by invoking a known formation mechanism of tight binaries under the influence of a misaligned third star and Kozai-Lidov cycles. The applicability of this story has been called into question lately by \citet{moe2018}, who deduce that Kozai-Lidov is only responsibility for the minority of tight binary formation. More theoretical work is needed. Additionally, \citet{munoz2015,martin2015a,hamers2016} suggested that any planets found orbiting around very tight binaries would be likely small and/or misaligned, both of which have been difficult to find to date. 
\item \textbf{\textit {There is an over-abundance of planets orbiting near the dynamical stability limit.}} This is likely the result of migratory formation of the planets, stalling near the edge of an inner disc cavity, which roughly coincides with the dynamical stability limit \citep{holman1999,kley2019}. Whilst this is not the sole result of an observational bias (Fig.~\ref{fig:period_semi_ratios}), more detections are needed to determine its statistical significance \citep{martin2014,li2016}. In particular, finding circumbinary planets by radial velocities \citep{martin2019a} or microlensing \citep{bennett2016} would allow for planet detections at longer periods, farther from the stability limit. 
\item \textbf{\textit {All transiting circumbinary planets are larger than $3R_{\oplus}$}}. This is contrary to the abundant discoveries of small planets around single stars, with a comparison shown in Fig.~\ref{fig:period_radius}. If it were a real absense, it would be enlightening, however the lack of small circumbinary planets is a detection bias; the days-amplitude transit timing variations (TTVs, \citealt{armstrong2013}) inhibit traditional planet detection techniques based on phase-folding on a fixed period. Only transits of giant planets could be found so far, by eye.  Some algorithms have been proposed to find small circumbinary planets, using modified versions of Boxed Least Squares (BLS, \citealt{ofir2008}) and the Quasiperiodic Automated Transit Search (QATS, \citealt{windemuth2019b}), but no new candidates have been reported yet.
\end{enumerate}

\section{A Search For Small Transiting Circumbinary Planets In Kepler}

\begin{figure*}  
\begin{center}  

\includegraphics[width=\textwidth]{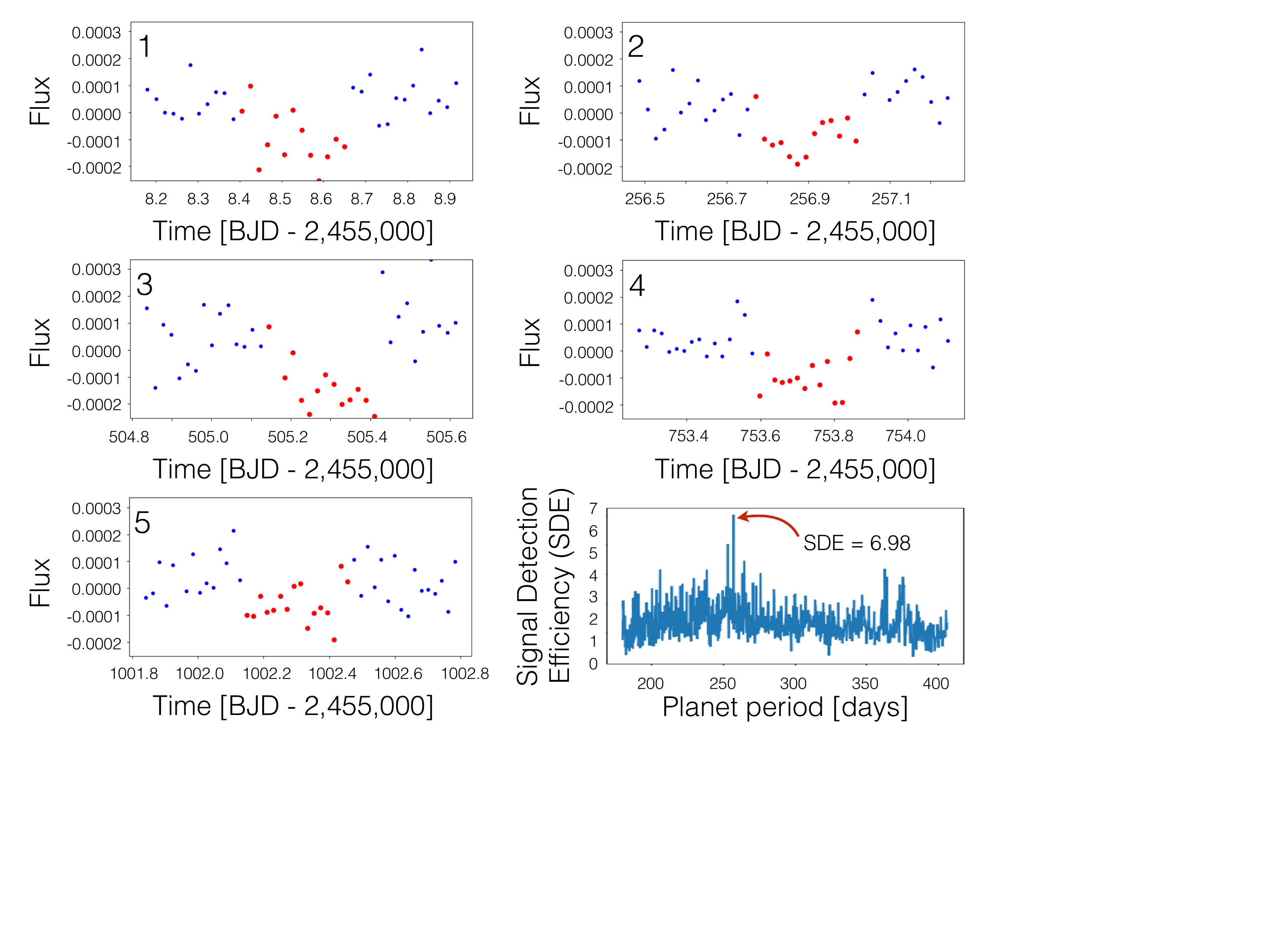}  

\caption{Recovery of an injected 260-day $0.875R_{\oplus}$ circumbinary planet on Kepler-16 (real planet is over $8R_{\oplus}$ with a 228 day period) with the new automated algorithm. Injected transits were created using {\it BATMAN} \citep{kreidberg2015}, with the duration scaled according to the relative planet-star velocity calculated by the {\it REBOUND} N-body algorithm \citep{rein2012}}.

\label{fig:injection}
\end{center}  
\end{figure*}

The archival Kepler data remains the best source for finding small circumbinary planets, because of its long four-year baseline, high-precision photometry and well-characterised EB catalog \citep{prsa2011,windemuth2019b}. In collaboration with Dan Fabrycky, a new transit search algorithm is being specifically designed for shallow transits of small circumbinary planets. It can successfully recover all known circumbinary planets, and also injected planets  slightly smaller than Earth (Fig~\ref{fig:injection}). Planet detection is assisted by a detrending algorithm designed specific to EBs, which accounts for the variable length of circumbinary planet transits as a function of the binary phase. Unique to this transit detection algorithm is building TTVs directly and exactly into the search. For each set of orbital parameters the algorithm produces a quasi-periodic mask of transit times and durations using a rapid N-body algorithm. This mask is matched to the photometric data similar to the cross correlation technique for high-precision RV fits to spectroscopic data. The N-body-derived mask fully incorporates the three-body geometry and both short and long-term dynamical variations of the planet's orbit. The search grid has been optimised using principles similar to \citep{ofir2014}, but adapted to circumbinary planets. 

Roughly two dozen detections are expected if planets have a similar size distribution around one and two stars (preliminary research suggests this is the case for gas giants, \citealt{martin2014,armstrong2014}). Alternatively, it is possible that small circumbinary planets \textbf{\textit {are rare or non-existent}}. This would suggest that super-Earths form in situ rather than with significant migration, helping answer a hotly-debated topic \citep{ogihara2015}; around single stars such a process is possible but around a binary it would be suppressed \citep{paardekooper2012}.

\section{The BEBOP radial velocity survey}

Between 2013 and 2018 a blind survey for circumbinary planets was run on the Swiss Euler Telescope. It was given the delightful name BEBOP - ``Binaries Escorted By Orbiting Planets''. BEBOP uniquely targetted eclipsing, single-lined spectroscopic binaries. The eclipses add preferential biases in both radial velocity amplitude and transit probability \citep{martin2015b,martin2017}. The single-lined binaries, composed of F/G primaries and M-dwarf secondaries, avoid the difficult problem of spectral contamination, and the need to deconvolve two moving sets of spectral lines. This is different to the SB2 search of  TATOOINE \citep{konacki2009}.

\begin{figure*}  
\begin{center}  
\includegraphics[width=\textwidth]{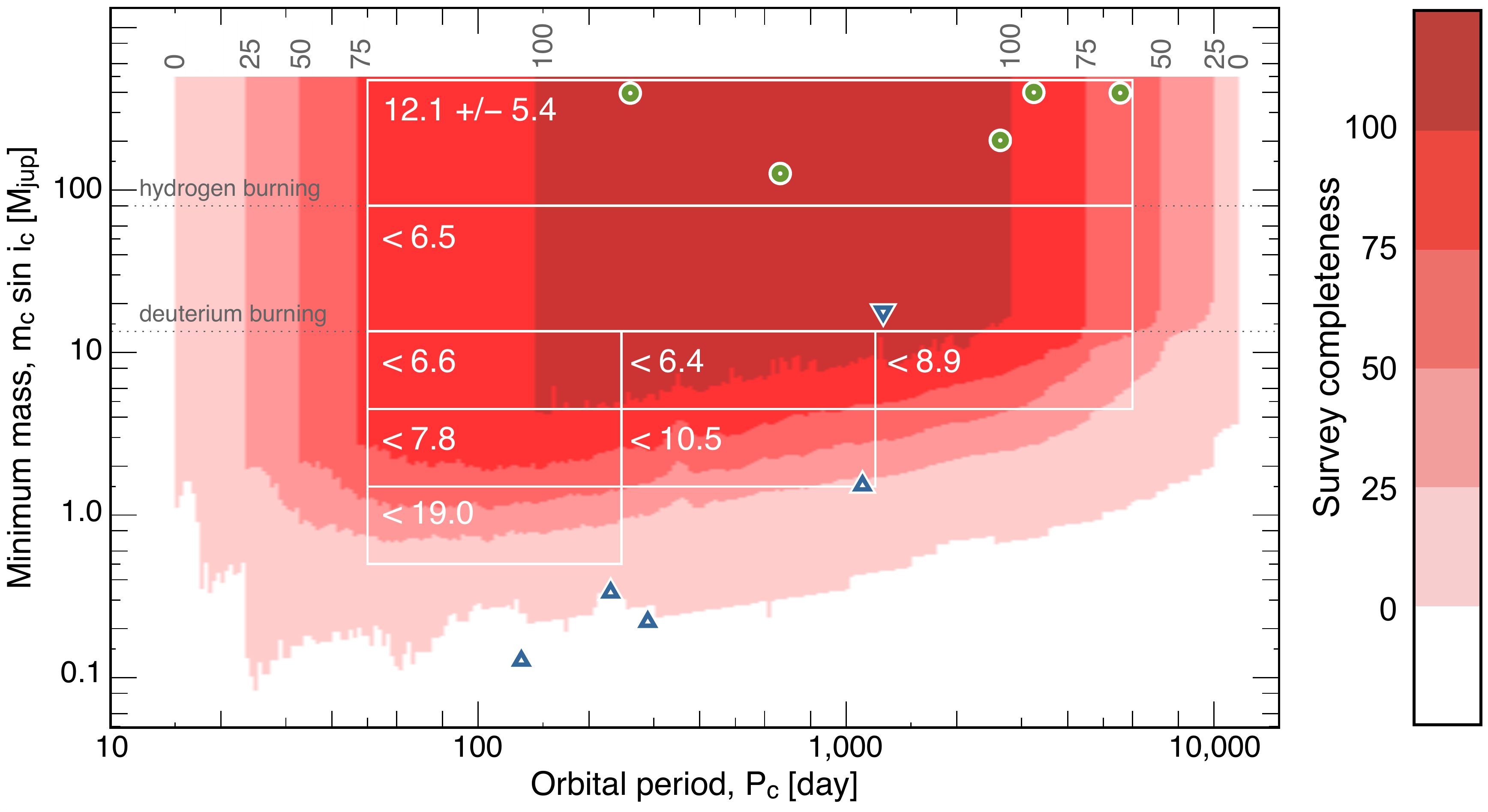}  

\caption{BEBOP detection completeness, detected triple star systems (green circles), known transiting circumbinary planets with roughly-characterised masses (upwards blue triangles) and a known circumbinary brown dwarf (downards blue triangle). Numbers in white boxes indicate $95\%$ confidence abundance bounds. Figure reproduced from \citet{martin2019a}.}

\label{fig:bebop}
\end{center}  
\end{figure*}

Over 1000 observations taken over more than 60 nights were compiled in \citet{martin2019a}. The survey was sensitive down to $0.5M_{\rm Jup}$, but our lack of detections showed that circumbinary planets are typically sub-Saturn mass (Fig.~\ref{fig:bebop}). BEBOP was sensitive to planetary mass companions at periods of several years, much longer than the Kepler discoveries. BEBOP also demonstrated that there was not a large abundance of giant, misaligned planets, which were proposed by \citet{martin2014,armstrong2014} as compatible with the Kepler transit results. BEBOP has since been expanded to large programs on HARPS, SOPHIE and ESPRESSO.

%

 \begin{figure*}  
\begin{center}  

\includegraphics[width=\textwidth]{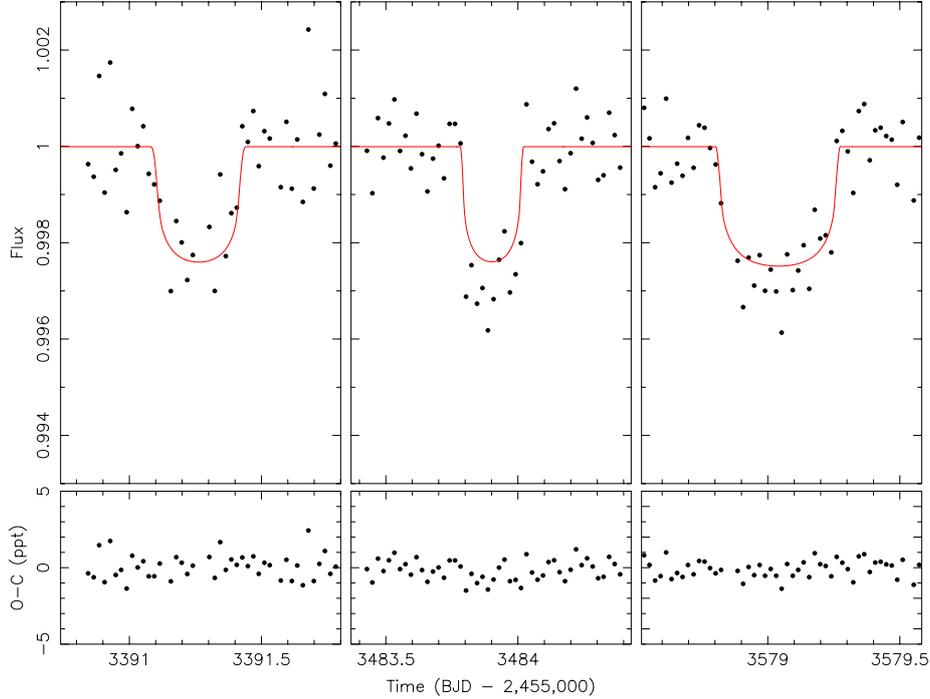}  

\caption{Three primary transits of the circumbinary planet TOI-1338/EBLM J0608-59  and the photodynamical fit with its residuals (observed minus calculated). The variable transit duration, owing to a variable relative velocity between the star and planet, is a smoking-gun signature of a circumbinary planet. Figure reproduced from Kostov et al. (under review).}

\label{fig:tess_transits}
\end{center}  
\end{figure*} 

\section{The TESS transit mission}

TESS presents different challenges and opportunities when compared with Kepler. TESS is observing most of the sky, in both hemispheres, and hence is targeting many more bright stars so ground-based follow-up is significantly easier. However, two drawbacks are the smaller telescope size (10.5 cm compared with 95 cm) and shorter observing timespans (30 days for most of the TESS field). Only near the ecliptic poles does the TESS timespan increase to almost a year of continuous viewing, owing to the overlap of multiple sectors. Indeed, the single TESS planet found so far is near the continuous viewing zone: TOI-1338/EBLM J0608-59 (Kostov et al. under review, Fig.~\ref{fig:tess_transits}). The planet has very similar properties to the Kepler population of planets (it is highlighted in Fig~\ref{fig:trends}). A unique aspect of this discovery, compared with the Kepler discoveries, is that the binary was already known and well characterised as a part of the EBLM \citep{triaud2017} and BEBOP \citep{martin2019a} radial velocity surveys, and those measurements were vital to the planet's characterisation.

TESS is unlikely to significantly break into new parameter spaces of circumbinary planets, due to the shortened observational timespans and inferior photometric precision to Kepler. Most detections will be harder than TOI-1338/EBLM J0608-59. What TESS will hopefully provide though is a significant increase in the statistics of circumbinary planets. The TESS circumbinary planet working group predicts 140 TESS circumbinary planets if we can detect them on a single passing that transits both stars, a ``1-2 punch''. This is based on 400,000 eclipsing binaries, a Kepler-like circumbinary planet detection rate of 11/2800, a 30/180 chance of the median circumbinary period transiting during a one month window and a 1/2 chance of the planet transiting both stars \citep{martin2017}. Based on Kepler, $\sim50$ of these planets are expected to be in the habitable zone.

\bibliography{martinDavid}
\end{document}